\documentclass[reprint, prb, superscriptaddress, dvipdfmx]{revtex4-1}
\usepackage{amsmath, graphicx}
\usepackage{newtxtext, newtxmath}
\usepackage{color}

\newcommand{\ie}{i.e. }

\newcommand{\e  }{\mathrm{e}}
\renewcommand{\d}{\mathrm{d}}
\newcommand{\iu }{i}

\newcommand{\cm}{\mathrm{cm^{-1}}}
\newcommand{\fs}{\text{fs}}
\newcommand{\ps}{\text{ps}}
\newcommand{\K }{\text{K}}

\newcommand{\A}{\mathrm{A}}
\newcommand{\D}{\mathrm{D}}
\newcommand{\g}{\mathrm{g}}

\newcommand{\chl }{\text{Chl}}
\newcommand{\pheo}{\text{Pheo}}
\newcommand{\Qy  }{\mathrm{Q_y}}
\newcommand{\vib }{\text{vib}}

\renewcommand{\H   }{H}
\newcommand{\X     }{X}
\newcommand{\sys   }{\text{sys}}
\newcommand{\env   }{\text{env}}
\newcommand{\sysenv}{\text{sys-env}}
\newcommand{\calE  }{\mathcal{E}}

\newcommand{\eq}{eq}
\newcommand{\eqs}{eqs}
\newcommand{\fig}{Figure}
\newcommand{\figs}{Figures}

\newcommand{\refers}{refs}
\newcommand{\refer}{ref}

\begin{document}
% Title %%%%%%%%%%%%%%%%%%%%%%
\title{Intramolecular Vibrations Complement the Robustness of Primary Charge Separation in a Dimer Model of the Photosystem II Reaction Center}
% Authors %%%%%%%%%%%%%%%%%%%%
\author{Yuta Fujihashi}
\affiliation{Institute for Molecular Science, National Institutes of Natural Sciences, Okazaki 444-8585, Japan}
\author{Masahiro Higashi}
\affiliation{Department of Chemistry, Biology, and Marine Science, University of the Ryukyus, 1 Senbaru, Nishihara, Okinawa 903-0213, Japan}
\author{Akihito Ishizaki}
\email{ishizaki@ims.ac.jp}
\affiliation{Institute for Molecular Science, National Institutes of Natural Sciences, Okazaki 444-8585, Japan}
\affiliation{School of Physical Sciences, The Graduate University for Advanced Studies, Okazaki 444-8585, Japan}
%
%\date{\today}
%\begin{document}
%\begin{tocentry}
%\includegraphics{TOC.pdf}
%\end{tocentry}
% Abstract %%%%%%%%%%%%%%%%%%%
\begin{abstract}
The energy conversion of oxygenic photosynthesis is triggered by primary charge separation in proteins at the photosystem II reaction center. Here, we investigate the impacts of the protein environment and intramolecular vibrations on primary charge separation at the photosystem II reaction center. This is accomplished by combining the quantum dynamic theories of condensed phase electron transfer with quantum chemical calculations to evaluate the vibrational Huang-Rhys factors of chlorophyll and pheophytin molecules. 
We report that individual vibrational modes play a minor role in promoting the charge separation, contrary to the discussion in recent publications. Nevertheless, these small contributions accumulate to considerably influence the charge separation rate, resulting in sub-picosecond charge separation almost independent of the driving force and temperature.
We suggest that the intramolecular vibrations complement the robustness of the charge separation in the photosystem II reaction center against the inherently large static disorder of the involved electronic energies.
\end{abstract}
\maketitle

% Introduction %%%%%%%%%%%%%%%
Oxygenic photosynthesis in plants, cyanobacteria, and algae begins in photosystem II (PSII).\cite{Renger:2008cr, vanAmerongen:2013ep}
The reaction center (RC) of PSII contains six chlorophyll (Chl) and two pheophytin (Pheo) molecules arranged in the form of two nearly symmetric branches corresponding to the D1 and D2 proteins. The PSII RC and the well-investigated purple bacterial RC share considerable similarity in the arrangement of their redox cofactors,\cite{Renger:2010ko} and thus it was speculated that the manner of primary charge separation in the PSII RC would be similar to that of purple bacteria. 
In the last two decades, however, it was recognized that the charge separation in the PSII RC most likely proceeded in a manner that is different from that at the purple bacterial RC.\cite{Durrant:1995jn, Groot:1997tj, Prokhorenko:2000hd, Romero:2017fc}
The nature of the charge separation in the PSII RC was investigated with the use of femtosecond pump-probe spectroscopy in the visible/mid-infrared \cite{Groot:2005tk} and visible \cite{Holzwarth:2006wt} spectral regions. 
Both reports identified the accessory chlorophyll ($\chl_{\rm D1}$) as the primary electron donor and pheophytin ($\pheo_{\rm D1}$) as the primary acceptor.
Time constants of $600-800\,\fs$ \cite{Groot:2005tk} and $3\,\ps$ \cite{Holzwarth:2006wt} were extracted for the pheophytin reduction, yielding $200-300\,\fs$ and $1\,\ps$ as the intrinsic time constant of the primary charge separation.\cite{Renger:2008cr} 
Moreover, theoretical analyses of time-dependent emission from the PSII core complex yielded $100\,\fs$ as the intrinsic time constant. \cite{Raszewski:2008go}
Regardless of the controversial differences, all values for the PSII RC are one-order magnitude faster than the time constant measured for the charge separation starting from the special pair in purple bacterial RCs. \cite{Fleming:1988hk} As the coupling strengths between electron donors and acceptors are usually thought to be of the order of tens of wavenumbers,\cite{Fleming:1988hk, Novoderezhkin:2007cl, Wang:2007it} the precise mechanisms that enable sub-picosecond charge separation are to a large extent unknown.

\begin{figure}%%%%%%%%%%%%%%%%%%%%
	\includegraphics{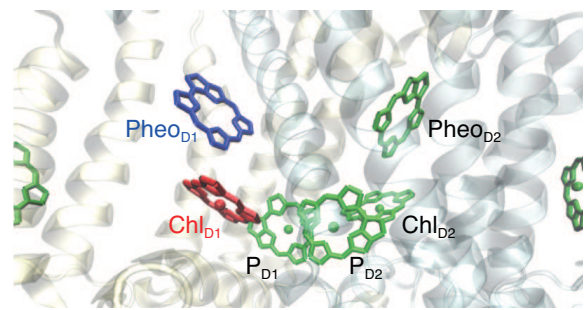}
	\caption{The structural arrangement of pigments in the PSII reaction center. 
   Data taken from the 4UB6 PDB structure.\cite{Suga:2015ho} The tails of chlorophylls and pheophytins are removed for the sake of clarity.}
	\label{fig:PSIIRC}
\end{figure}%%%%%%%%%%%%%%%%%%%%%%

Recently, Romero et al.\cite{Romero:2014jm} and Fuller et al.\cite{Fuller:2014iz} revealed the presence of long-lived quantum beats in the PSII RC by means of two-dimensional (2D) electronic spectroscopy.\cite{SchlauCohen:2011cw, Scholes:2017gw}
Many of the observed beats possess frequencies of vibrational modes identified in resonance Raman \cite{Picorel:1998jw} and fluorescence line-narrowing spectra,\cite{Peterman:1998fo} and some of the frequencies were deemed to match the frequency differences between electronic excitons and the primary charge transfer state. 
On the basis of the experimental observations, the authors suggested that the electronic-vibrational resonance might represent an important design principle for enabling charge separation with high quantum efficiency in oxygenic photosynthesis.
Namely, resonance between an electronic exciton state and the vibrational levels in the charge-transfer state leads to quantum mechanically mixed electronic and vibrational states, and thereby optimizes the flow of electrons to the final charge-separated state.
However, the reorganization energies and thus the protein-induced fluctuations associated with charge transfer states are generally large,\cite{Renger:2004cz, Mancal:2006gf, Gelzinis:2013ch, Gelzinis:2017cz} and hence it is questionable whether such electronic-vibrational mixtures could be robust and could play a role under the influence of the fluctuations at physiological temperatures.\cite{Ishizaki:2010ft}
Indeed, by employing numerically accurate quantum dynamics calculations, Fujihashi et al.\cite{Fujihashi:2015kz} and Monahan et al.\cite{Monahan:2015gu} demonstrated that such electronic-vibrational mixtures do not necessarily play a role in hastening electronic energy transfer in protein environments, despite contributing to the enhancement of long-lived quantum beating in 2D electronic spectra at cryogenic temperatures. 
Furthermore, it should be noted that the static disorder in systems with larger reorganization energy is stronger. Indeed, Gelzinis et al.\cite{Gelzinis:2017cz} revealed strong static disorder ($\sim 550\,\cm$) by analyzing multiple optical spectra of the PSII RC. Hence, further investigations on the influence of intramolecular vibrations on primary charge separation are required.

In this study, we comprehensively investigate the impacts of the intramolecular vibrational modes on the primary charge separation starting from the accessory Chl in the PSII RC by combining quantum dynamic theories of condensed phase electron transfer with quantum chemical calculations for evaluating the vibrational Huang-Rhys factors in Chl and Pheo molecules as well as the parameters extracted from experimental measurements.

% Theory %%%%%%%%%%%%%%%
For the sake of simplicity, we do not explicitly consider exciton-charge-transfer states such as $( \chl^{\delta +} \pheo^{\delta -} )^\ast $, where $\delta\pm$ indicates charge-transfer character.\cite{Romero:2017fc, Romero:2014jm, Romero:2012ks}
Instead, we consider a simpler scheme for the primary charge separation process,
\[
	\chl + \pheo 
	\xrightarrow{h\nu}
	\chl^\ast + \pheo
	\to 
	\chl^+ +\pheo^-.
\]
It should be noticed that this scheme does not contradict the possible existence of the exciton-charge-transfer states. Fuller discussion on the exciton-charge-transfer states will be presented later in conjunction with \eqs~\ref{eq:employed-eq} and \ref{eq:free_enegy_activation}.
This reaction involves three states: the electronic ground state $\lvert \g \rangle$, the photo-excited electron-donor state $\lvert \D \rangle$, and the electron-acceptor state $\lvert \A \rangle$. The dynamics of the charge separation is described by the Hamiltonian,
\begin{align}
	\H
	=
	\sum_{m=\g, \D, \A} \H_m \lvert m \rangle\langle m \rvert  
	+ V_{\D\A} (\lvert \D \rangle\langle \A \rvert + \lvert \A \rangle\langle \D \rvert),
	\label{eq:primary-CS-hamiltonian}
\end{align}
with $\H_\g = \H_\chl + \H_\pheo$, $\H_\D = \H_{\chl^\ast} + \H_\pheo$, and $\H_\A = \H_{\chl^+} + \H_{\pheo^-}$, where $\H_\chl$, $\H_{\chl^\ast}$, $\H_{\chl^+}$, $\H_\pheo$ and $\H_{\pheo^-}$ represent the diabatic Hamiltonians to describe the intramolecular vibrational modes of the respective molecular states and the associated environmental degrees of freedom (DOFs). The interstate coupling, $V_{\D\A}$, is assumed to be independent of the accessible environmental and vibrational DOFs.

\begin{figure}%%%%%%%%%%%%%%%%%%%%
\includegraphics{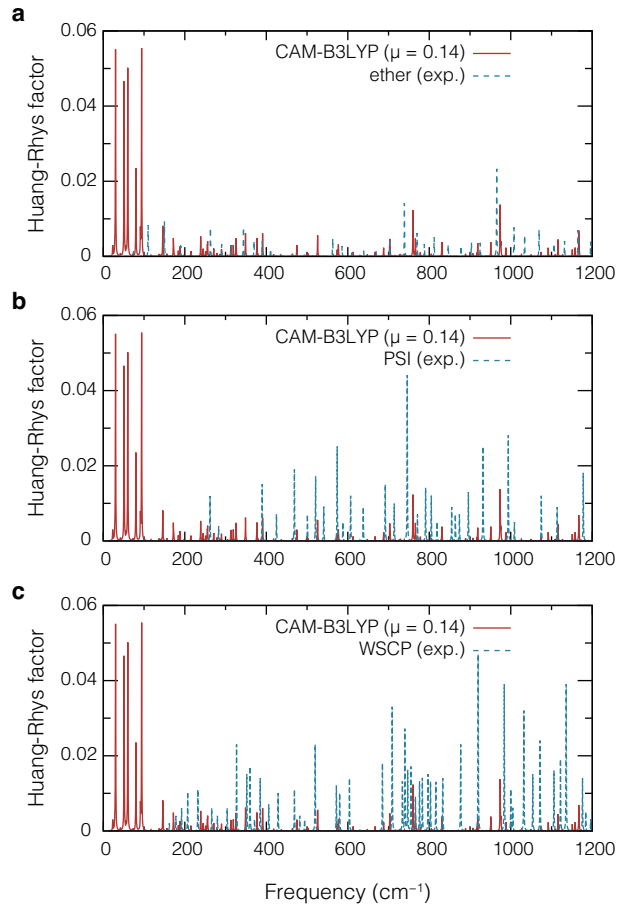}
\caption{Calculated Huang-Rhys factors of intramolecular vibrational modes associated with electronic excitation of chlorophyll\,\textit{a}.
	Calculations were performed with the use of time-dependent density functional theory, CAM-B3LYP/6-31G(d)\cite{Yanai:2004ia} with $\mu=0.14$. For reference purposes, the experimentally evaluated Huang-Rhys factors are also presented:
	(a) high-resolution fluorescence excitation spectrum of Chl{\it a} in ether at 4.2\,K,\cite{Reimers:2013fy}
	(b) hole burning experiment on photosystem I at 1.6\,K,\cite{Gillie:1989gt, Reimers:2013fy}
	(c) hole burning experiment on the water-soluble chlorophyll-binding protein at 1.4\,K.\cite{Hughes:2010co, Reimers:2013fy}}
	\label{fig:HR-factors-S0S1}
\end{figure}

To evaluate the Huang-Rhys factors of the intramolecular vibrational modes for the transitions, $\chl \to \chl^\ast$, $\chl^\ast \to \chl^+$, and $\pheo \to \pheo^-$, we performed electronic structure calculations with both the Gaussian 16\cite{g16} and DUSHIN programs.\cite{Reimers:2001jt} 
The fully optimized Cartesian displacements between the two adiabatic potential energy minima are projected onto the normal modes by using the DUSHIN program, and thereby the dimensionless normal mode displacements and the corresponding Huang-Rhys factors are obtained.
Technical details of the calculations are given in Methods.
Figure~\ref{fig:HR-factors-S0S1} presents the Huang-Rhys factors for the $\Qy$ transition of Chl{\it a} calculated with the use of time-dependent density functional theory, CAM-B3LYP/6-31G(d) \cite{Yanai:2004ia} with $\mu=0.14$.
For reference purposes, the experimentally evaluated Huang-Rhys factors are also presented.
Figure~\ref{fig:HR-factors-S0S1}a shows the Huang-Rhys factors obtained from the high-resolution fluorescence excitation spectrum of Chl\textit{a} in ether at 4.2\,K.\cite{Reimers:2013fy} The calculated and experimental results are in reasonably good agreement, except for the low-frequency modes, $\omega_\xi < 150\,\cm$. 
The large discrepancy in the low-frequency region may be attributed to the harmonic approximation of the low-frequency vibrational modes in the DUSHIN program. Typically, low-frequency modes exhibit strong anharmonicity;\cite{Ribeiro:2011kq} however, the DUSHIN program\cite{Reimers:2001jt} maps all of the modes onto harmonic normal modes. In what follows, the low-frequency modes ($\omega_\xi < 150\,\cm$) are excluded.
Figures~\ref{fig:HR-factors-S0S1}b and \ref{fig:HR-factors-S0S1}c show the Huang-Rhys factors evaluated by means of a hole-burning experiment on photosystem I isolated from the chloroplast of spinach at 1.6\,K,\cite{Gillie:1989gt} and the water-soluble chlorophyll-binding protein in cauliflower at 1.4\,K,\cite{Hughes:2010co} respectively.
The Huang-Rhys factors for the $\Qy$ transition of Chl{\it a} embedded in various photosynthetic proteins differ significantly from each other, indicating significant dependence of the Huang-Rhys factors and the vibrational distribution on local environments. 
Figures~\ref{fig:HR-factors-S0S1}b and \ref{fig:HR-factors-S0S1}c also demonstrate that protein environments may increase the Huang-Rhys factors by several times compared with the calculated ones.

%%%%%%%%%%%%%%%%%%%%%%

\begin{figure}%%%%%%%%%%%%%%%%%%%%
\includegraphics{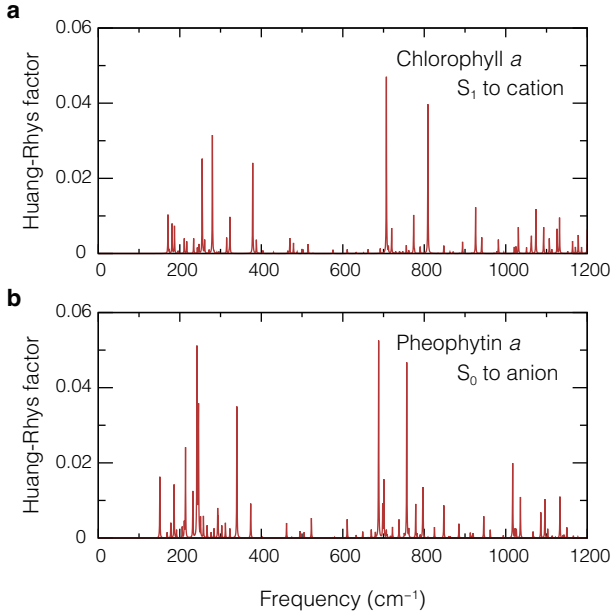}
\caption{Calculated Huang-Rhys factors of intramolecular vibrational modes associated with transitions involved in the charge separation.
	(a) $\chl a^\ast \to \chl a^+$ and (b) $\pheo a \to \pheo a^-$. The calculations were performed with the use of time-dependent density functional theory, CAM-B3LYP/6-31G(d) \cite{Yanai:2004ia} with $\mu=0.14$.}
\label{fig:HR-factors-CS}
\end{figure}%%%%%%%%%%%%%%%%%%%%%%

\begin{figure}%%%%%%%%%%%%%%%%%%%%
\includegraphics{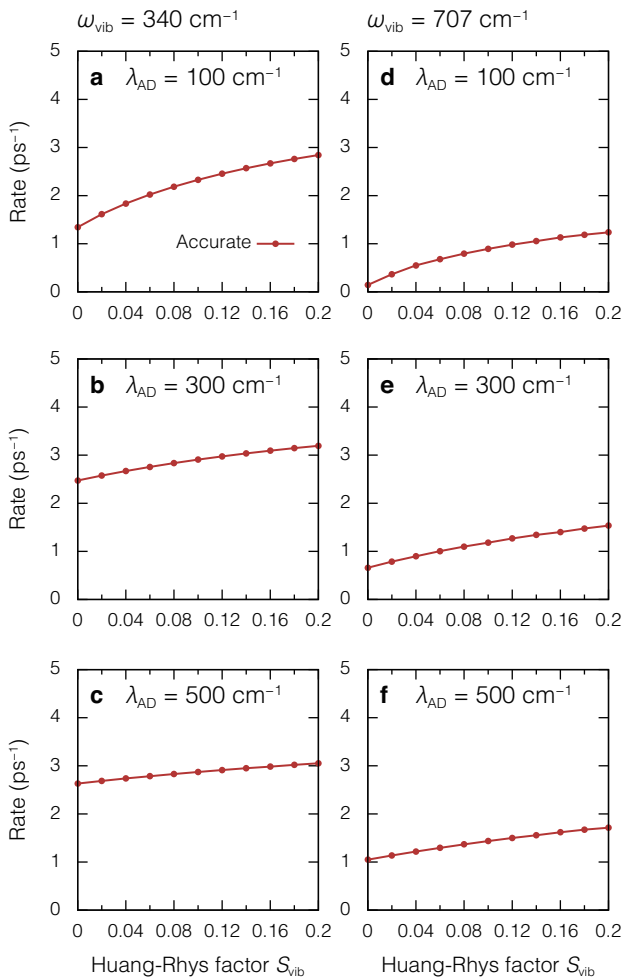}
\caption{Huang-Rhys factor dependence of the charge separation rates.
	The calculated rates are presented as a function of the Huang-Rhys factor $S_\vib$ for the case of $-\Delta G^\circ = \hbar\omega_\vib + \lambda_{\A\D}$, where the vibrationally excited state in the acceptor state resonates with the equilibrated donor state.
	The rates were obtained with the use of the numerically accurate quantum dynamics calculations. 
	The panels on the left ({a}--{c}) present the results for the vibrational mode of frequency $\omega_\vib=340\,\cm$, whereas those on the right ({d}--{f}) contain the results for the mode of $\omega_\vib=707\,\cm$.
	The other parameters are set to be $V_{\D\A}=70\,\cm$, $\tau=50\,\fs$, $\gamma_\vib^{-1}=2\,\ps$, and $T=300\,\K$.}
\label{fig:rate-S-dep}
\end{figure}

In contrast to the $\Qy$ transition, the Huang-Rhys factors associated with the charge separation, $\chl^\ast + \pheo \to \chl^+ + \pheo^-$, are inaccessible in a spectroscopic fashion, and \fig~\ref{fig:HR-factors-CS} presents the calculated Huang-Rhys factors for the transitions of (a) $\chl^\ast \to {\rm Chl}^+$ and (b) $\pheo \to \pheo^-$. 
The main features are consistent with the frequencies of the dominant vibrational modes extracted from the beating of the 2D electronic spectra of the PSII RC ($250\, \cm$, $340\,\cm$, and  $730\,\cm$).\cite{Romero:2014jm,Fuller:2014iz}
However, it should be noticed that the protein environment may increase the Huang-Rhys factors, as was discussed in \fig~\ref{fig:HR-factors-S0S1}. This issue will be taken into account in discussing impacts of the intramolecular vibrational modes on the primary charge separation in the PSII RC (\fig~\ref{fig:rate-S-dep}).

\begin{figure}%%%%%%%%%%%%%%%%%%%%
\includegraphics{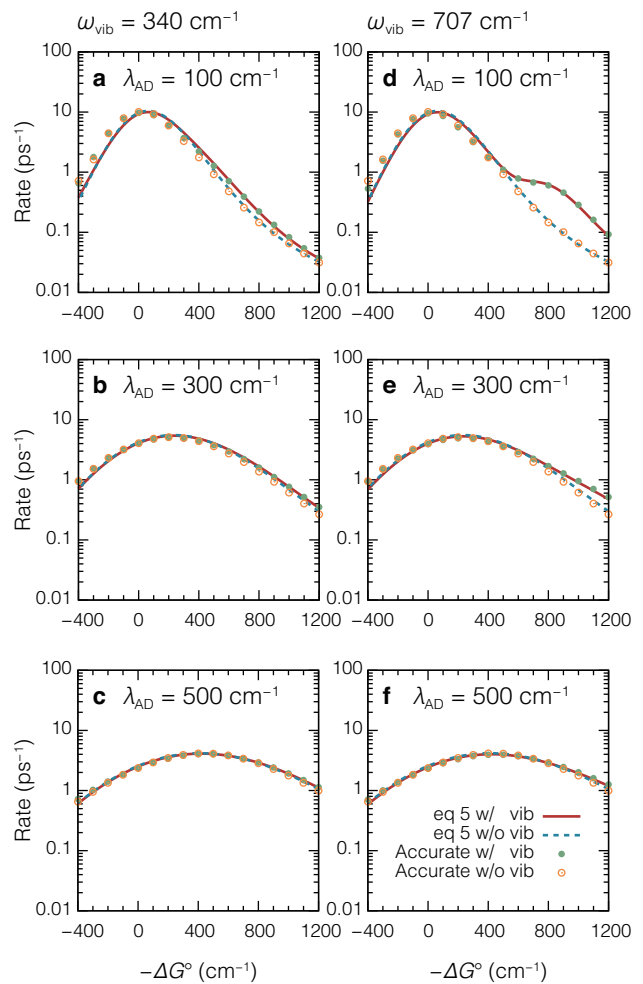}
\caption{Rates of the charge separation influenced by specific vibrational modes, $\rm 340\,\cm$ and $\rm 707\,\cm$.
	The calculated rates are presented as a function of the driving force $-\Delta G^\circ=E_\D^\circ - E_\A^\circ$ for various values of the environmental reorganization energy associated with the charge separation, $\lambda_{\A\D}$.
	Calculations were conducted with \eq~\ref{eq:employed-eq} with $\alpha=0.005/\hbar$ (red solid lines) and the numerically accurate quantum dynamics calculation (green filled circles) including a vibrational mode.
	For the purpose of comparison, results in the absence of the vibrational mode are also shown (blue dashed lines, orange open circles).
	The panels on the left ({a}--{c}) present the results for the vibrational mode of frequency $\omega_\vib=340\,\cm$ and the Huang-Rhys factor $S_\vib = 0.035$, whereas those on the right ({d}--{f}) for the mode of $\omega_\vib=707\,\cm$ and $S_\vib = 0.047$.
	The other parameters are set to be $V_{\D\A}=70\,\cm$, $\tau=50\,\fs$, $\gamma_\vib^{-1}=2\,\ps$, and $T=300\,\K$.}
\label{fig:rate-1mode}
\end{figure}

The non-equilibrium reorganization process of the environment may strongly influence the electron transfer reaction in the case of the large amount of reorganization energy associated with the photoexcitation.\cite{Cho:1995kf, Ishizaki:2013jg}
However, the environmental reorganization energy and the vibrational Huang-Rhys factors associated with the photoexcitation of chlorophyll in the visible region are small,\cite{Ratsep:2007fq, Ratsep:2011cq, Reimers:2013fy} and the interstate coupling $V_{\D\A}$ is typically tens of wavenumbers.\cite{Novoderezhkin:2011en,Gelzinis:2013ch} 
Hence, it may be assumed that the environmental reorganization is completed and the environmental and vibrational DOFs are equilibrated prior to the charge separation. 
Although excited vibrational levels in the donor state are slightly populated and contribute to the quantum beats in 2D electronic spectra, such vibrational excitations are of no consequence with regard to the electronic-vibrational mixing. As was demonstrated in \refer~\citenum{Ishizaki:2013jg}, furthermore, the Marcus-type thermal electron transfer could be an appropriate description for typical time constant ($\sim 50\,\fs$) of the non-equilibrium environmental reorganization associated with the photo-excited donor state.
In this situation, the rate of charge separation can be given by the second-order perturbative truncation in terms of the interstate coupling $V_{\D\A}$:
\begin{align}
	k^{(2)} 
	=
	\frac{2 {V_{\D\A}}^2}{\hbar^2}
	\mathrm{Re}
	\int^\infty_0 \d t\,
	\langle \e^{ \iu \H_\D t/\hbar} \e^{-\iu \H_\A t/\hbar} \rangle_\D.
	\label{eq:FGR-rate}
\end{align}
In the equation, $\langle \dots \rangle_m$ denotes the statistical average with respect to the equilibrium states of the environmental and vibrational DOFs associated with the state $\lvert m \rangle$, $\rho^{\rm eq}_m=\e^{-\beta\H_m}/\mathrm{Tr}\,\e^{-\beta\H_m}$, where $\beta$ represents the inverse temperature $\beta=1/k_{\rm B}T$ with the Boltzmann constant $k_{\rm B}$ and temperature $T$.
Here, we introduce the Franck-Condon vertical transition energy from the equilibrated electron donor state to the acceptor state, $\hbar\Omega_{\A\D}=\langle \H_\A - \H_\D \rangle_\D$, and the collective energy gap coordinate, \cite{Marchi:1993gu, Gehlen:1994ke} $\X_{\A\D} = \H_\A - \H_\D - \hbar\Omega_{\A\D}$,
which contains information on the fluctuations in the electronic energies of the electron donor and acceptor states and on the relevant nuclear dynamics.
In this work, we assume that the environmentally induced fluctuations can be described as Gaussian processes\cite{Marchi:1993gu} and that the relevant nuclear dynamics are harmonic. 
Under this assumption, \eq~\ref{eq:FGR-rate} is recast into
\begin{align}
	k^{(2)} 
	= 
	\frac{2{V_{\D\A}}^2}{\hbar^2} 
	\mathrm{Re}
	\int^\infty_0 \d t\, \exp[ -\iu \Omega_{\A\D}t - g_\D(t) ].
	\label{eq:cummurant-rate}
\end{align}
The Franck-Condon vertical transition energy $\hbar\Omega_{\A\D}$ is expressed as
\begin{align}
	\hbar\Omega_{\A\D}
	=
	\Delta G^\circ + \lambda_{\A\D} + \hbar\sum_{\xi=1}^N S_\xi \omega_\xi,
\label{eq:FC-relation}
\end{align}
where $\Delta G^\circ$, $\lambda_{\A\D}$, $\omega_\xi$, and $S_\xi$ represent the driving force $E_\A^\circ - E_\D^\circ$, the environmental reorganization energy associated with the charge separation, the frequency, and the Huang-Rhys factor of the $\xi$th vibrational mode, respectively.
The so-called line-broadening function $g_\D(t)$ is given as $g_\D(t)  =  \int^t_0\d s\int^s_0 \d s' \, C(s')/\hbar^2$, where the quantum correlation function $C(t) = \langle \X_{\A\D}(t) \X_{\A\D}(0) \rangle_\D$ is expressed with the spectral density $J(\omega)$, namely $C(t) = (\hbar/\pi) \int^\infty_0 \d \omega\, J(\omega) [ \coth({\beta\hbar\omega}/{2})\cos\omega t - \iu \sin\omega t]$.
It is noted that \eq~\ref{eq:cummurant-rate} yields the Marcus formula\cite{Marcus:1993kx} and the Jortner-Bixon formula\cite{Jortner:1988jr} to describe the condensed phase electron transfer reaction by applying further approximations.
To evaluate \eq~\ref{eq:cummurant-rate}, the spectral density is decomposed into the environmental and vibrational contributions, $J(\omega) = J_\env(\omega) +  J_\vib(\omega)$.
We investigate the timescales of the environmental dynamics affecting the electronic transition energies by modelling the environmental component with the Drude-Lorentz spectral density,\cite{Mukamel:1995us} $J_\env(\omega) = {2\lambda_{\A\D} \tau \omega}/{(\tau^2\omega^2+1)}$, where $\tau$ indicates the time constant of the environmental reorganization dynamics associated with the transition from the electron donor state to the acceptor state.
However, \eq~\ref{eq:cummurant-rate} is not capable of fully describing the environmental dynamics such as the dynamic solvent effect on the charge separation. Hence, we employ the %adjusted 
rate expression adjusted with the adiabatic correction,\cite{Rips:1987bt,Sparpaglione:1988eo}
\begin{align}
	k 
	= \frac{1}{1+ \alpha  V_{\D \A}^2 \tau   } 
	k^{(2)},
	\label{eq:employed-eq}
\end{align}
where $\alpha$ is determined by comparing the $\tau$- and $V_{\D\A}$-dependence of the rate with the numerically accurate ones, yielding $\alpha=0.005/\hbar$ in this work.
Equation~\ref{eq:employed-eq} recovers the nonadiabatic rate expression for vanishingly small values of $\tau$ and $V_{\D\A}$, whereas it produces a reasonable approximation for the adiabatic case where quantum mixing between the donor and acceptor states is strong enough.
The vibrational component, $J_\vib(\omega)$ is modelled with the multimode Brownian oscillator model, \cite{Mukamel:1995us} in which the relaxation rate of each mode is given by $\gamma_\xi$.
The applicability of \eq~\ref{eq:employed-eq} is verified by comparing the resultant rates with the numerically accurate ones (\fig~\ref{fig:rate-1mode}).

%%%%%%%%%%%%%%%%%%%%%%

We explore the impacts of specific intramolecular vibrations of frequency $\omega_\vib$ and Huang-Rhys factor $S_\vib$. Figure~\ref{fig:rate-1mode} presents the rates of the charge separation at the physiological temperature $T=300\,\K$ as a function of the driving force, $-\Delta G^\circ$, calculated with \eq~\ref{eq:employed-eq} (red solid lines) and the numerically accurate quantum dynamics calculations (green filled circles).
For the vibrational modes involved in this calculations, we choose two of the intramolecular modes with relatively large Huang-Rhys factors for the transitions of $\chl^\ast \to \chl^+$ and $\pheo \to \pheo^-$ in \fig~\ref{fig:HR-factors-CS}, \ie $\omega_\vib=340\,\cm$, $S_\vib = 0.035$ (\figs~\ref{fig:rate-1mode}a--\ref{fig:rate-1mode}c) and $\omega_\vib=707\,\cm$, $S_\vib = 0.047$ (\figs~\ref{fig:rate-1mode}d--\ref{fig:rate-1mode}f). These two modes may correspond to the modes addressed by Romero et al.\cite{Romero:2014jm} and Fuller et al.\cite{Fuller:2014iz}
As the interstate coupling strength, we assume $V_{\D\A}=70\,\cm$, which is the value that is employed in the literature pertaining to the PSII RC. \cite{Novoderezhkin:2011en,Gelzinis:2013ch} The environmental reorganization time is set to be $\tau=50\,\fs$, and the vibrational relaxation rate is $\gamma_\vib^{-1}=2\,\ps$.
For reference purposes, the charge separation rates calculated in the absence of the vibrational contributions are also presented (blue dashed lines and orange open circles).
Figures~\ref{fig:rate-1mode}a--\ref{fig:rate-1mode}c indicate that the $340\,\cm$ vibrational mode does not play a significant role over a wide range of the reorganization energy and the driving force under the influence of the protein environment.
On the other hand, \figs~\ref{fig:rate-1mode}d--\ref{fig:rate-1mode}f shows that the $707\,\cm$ vibrational mode contributes more strongly to the enhancement of the rate, in particular in the vicinity of $- \Delta G^\circ = \hbar \omega_\vib + \lambda_{\A\D}$, where the vibrationally excited state in the acceptor state resonates with the equilibrated donor state.
However, an increase in the reorganization energy results in a decrease in the vibrational contribution. The larger amplitude of the fluctuations eradicate the electronic-vibrational resonance, as was discussed in \refers~\citenum{Ishizaki:2010ft,Fujihashi:2015kz}.

%%%%%%%%%%%%%%%%%%%%%%

However, it should not be overlooked that the Huang-Rhys factors in the reaction center protein may be several times larger than the values given in \fig~\ref{fig:HR-factors-CS}, as was discussed before.
To further clarify the extent of the vibrational contributions, we examine the dependence of the rate on the Huang-Rhys factor in \fig~\ref{fig:rate-S-dep}. The driving force is fixed to be $-\Delta G^\circ = \hbar\omega_\vib + \lambda_{\A\D}$.
Figures~\ref{fig:rate-S-dep}a--\ref{fig:rate-S-dep}c show the results for the $340\,\cm$ mode, and \figs~\ref{fig:rate-S-dep}d--\ref{fig:rate-S-dep}f show those for the $707\,\cm$ mode in accordance with \fig~\ref{fig:rate-1mode}. 
When the reorganization energy is small ($\lambda_{\A\D}=100\,\cm$), the rate enhancement caused by the larger Huang-Rhys factors is prominent.
Nevertheless, when the reorganization energy is assigned values that are physically more reasonable ($\lambda_{\A\D}>300\,\cm$), the rate becomes less sensitive to the Huang-Rhys factors.

\begin{figure}%%%%%%%%%%%%%%%%%%%%
\includegraphics{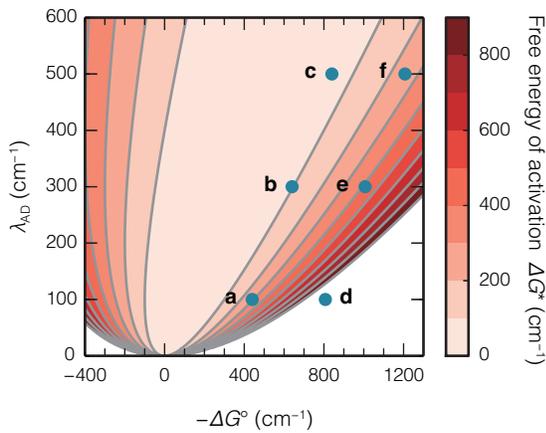}
\caption{Contour plot of the free energy of activation required for the charge separation to proceed.
	The free energy of activation $\Delta G^\ast$ is plotted as a function of the driving force $-\Delta G^\circ=E_\D^\circ-E_\A^\circ$ and the reorganization energy associated with the charge separation, $\lambda_{\A\D}$. Contour lines are drawn at $100\,\cm$ intervals.
	The marked points indicate the sets of $-\Delta G^\circ$ and $\lambda_{\A\D}$ that satisfy $-\Delta G^\circ = \hbar\omega_\vib + \lambda_{\A\D}$ for the vibrational frequencies $\omega_\vib=340\,\cm$ ({a}--{c}) and $\omega_\vib=707\,\cm$ ({d}--{f}), corresponding to \figs~\ref{fig:rate-1mode} and \ref{fig:rate-S-dep}.
	At physiological temperature $T=300\,\K$, the thermal energy is evaluated as $k_{\rm B}T \simeq 200\,\cm$.}
	\label{fig:barrier}
\end{figure}

To deepen our insight into the charge separation mechanism, the free energy surfaces of the electron donor and acceptor states are investigated with respect to the environmental component $\calE$ in the collective energy gap coordinate  $X_{\A\D}$,\cite{Marchi:1993gu, Gehlen:1994ke, Ishizaki:2013jg,Fujihashi:2016ig} 
\begin{align}
	G_\D(\calE) 
	&=
	E_\D^\circ + \frac{1}{4\lambda_{\A\D}}\calE^2,
	\\
	G_\A(\calE)
	&=
	E_\A^\circ + \frac{1}{4\lambda_{\A\D}}(\calE-2\lambda_{\A\D})^2.
	\label{eq:free_enegy_activation}
\end{align}
The intersection between $G_\D(\calE)$ and $G_\A(\calE)$ is located at $\calE^\ast=\Delta G^\circ + \lambda_{\A\D}$, yielding the free energy of activation that enables the charge separation to proceed as
\begin{align}
	\Delta G^\ast 
	\simeq 
	\frac{(\Delta G^\circ + \lambda_{\A\D})^2}{4\lambda_{\A\D}}.
\end{align} 
Figure~\ref{fig:barrier} presents 2D contour plots of $\Delta G^\ast$ as a function of the driving force $-\Delta G^\circ = E_\D^\circ - E_\A^\circ$ and the environmental reorganization energy $\lambda_{\A\D}$. Contour lines are drawn at $100\,\cm$ intervals.
The plots reveal that the free energy of activation is small in comparison with the thermal energy $k_{\rm B}T\simeq 200\,\cm$ at $T=300\,\K$ across a broad range of the 2D space. In such almost activationless situations, the primary charge separation takes place in a facile fashion without the help of high-frequency vibrational modes, and hence the vibrational modes play a minor role in promoting the charge separation.
In the situation satisfying $-\Delta G^\circ = \hbar\omega_\xi + \lambda_{\A\D}$, the first excited state of the $\xi$th mode in the acceptor state resonates with the equilibrated donor state, and thus the $\xi$th mode would be deemed to play a role in promoting the charge separation. In the situation corresponding to \fig~\ref{fig:rate-1mode}a ($-\Delta G^\circ = 440\,\cm$, $\omega_\xi=340\,\cm$ and $\lambda_{\A\D}=100\,\cm$), however, the free energy of activation is evaluated as  $289\,\cm$, which is comparable to the thermal energy, and hence the $340\,\cm$ vibrational mode plays a minor role in \fig~\ref{fig:rate-1mode}a. 
In the situation of \fig~\ref{fig:rate-1mode}d ($-\Delta G^\circ = 807\,\cm$, $\omega_\xi=707\,\cm$ and $\lambda_{\A\D}=100\,\cm$), on the other hand, the free energy of activation is $1249\,\cm$, which is much higher than the thermal energy. Therefore, the $707\,\cm$ mode plays a crucial role in enhancing the charge separation rate in \fig~\ref{fig:rate-1mode}d. The vibrational contributions in \figs~\ref{fig:rate-1mode}b, \ref{fig:rate-1mode}c, \ref{fig:rate-1mode}e, and \ref{fig:rate-1mode}f can be understood in a similar fashion.
Here, we note that a similar but different analysis was done by Abramavicius and Valkunas.\cite{Abramavicius:2015jy} They obtained an expression of the activation energy as a function of a high-frequency vibrational coordinate in addition to the driving force and the environmental reorganization energy, thereby discussing contributions of the high-frequency mode mainly in the normal region,  where vibrations in the donor state are effective. 
Ando and Sumi\cite{Ando:1998cs} and Novoderezhkin et al.\cite{Novoderezhkin:2004dh} also discussed coherent nuclear dynamics in the similar situation with regard to the primary charge separation in the purple bacterial reaction center.
Although their approaches differ from the present study, Abramavicius and Valkunas concluded that effects of the frequency match in electron transfer reaction are not as strict as in electronic energy transfer processes.\cite{Abramavicius:2015jy}

In conjunction with \fig~\ref{fig:barrier}, we now return to the issue of the exciton-charge-transfer state\cite{Romero:2017fc, Romero:2014jm, Romero:2012ks} $(\chl^{\delta+}\pheo^{\delta-})^\ast$ which was postponed in \eq~\ref{eq:primary-CS-hamiltonian}. The free energy of activation smaller than the thermal energy indicates that the free energy curves of the donor state ($\chl^\ast+\pheo$) and the acceptor state ($\chl^++\pheo^-$) intersect each other in the vicinity of the equilibrated donor state. In addition, the adiabatic correction in \eq~\ref{eq:employed-eq} manifests a certain amount of quantum mixing between the donor state ($\chl^\ast+\pheo$) and the acceptor state ($\chl^++\pheo^-$). Therefore, it is reasonable to consider that the initial state of the charge separation possesses the charge-transfer character, as was detected with Stark spectroscopy.\cite{Romero:2012ks}

%%%%%%%%%%%%%%%%%%%%%%

\begin{figure}%%%%%%%%%%%%%%%%%%%%
\includegraphics{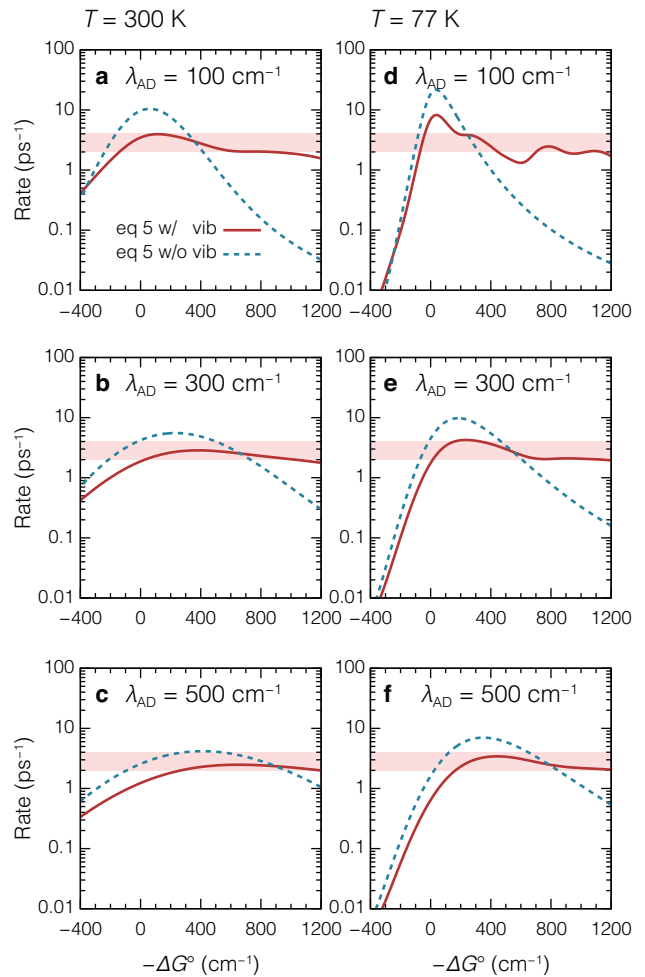}
\caption{Rate of the charge separation influenced by all of the vibrational modes in chlorophyll and pheophytin molecules.
	Calculated rates are presented as a function of the driving force $-\Delta G^\circ$ for various values of the environmental reorganization energy associated with the charge separation, $\lambda_{\A\D}$. Calculations were carried out with \eq~\ref{eq:employed-eq} with $\alpha=0.005/\hbar$ in the presence of all the vibrational contributions (red solid lines) and in the absence of the vibrational contribution (blue dashed lines). The panels on the left ({a}--{c}) present the results for the physiological temperature $T=300\,\K$, whereas those on the right ({d}--{f}) contain the results for the cryogenic temperature $T=77\,\K$.
	The Huang-Rhys factors presented in \fig~\ref{fig:HR-factors-CS} are employed to evaluate \eq~\ref{eq:employed-eq}. The other parameters are set to be $V_{\D\A}=70\,\cm$, $\tau=50\,\fs$, and $\gamma_\xi^{-1}=2\,\ps$.}
	\label{fig:rate-allmodes}
\end{figure}%%%%%%%%%%%%%%%%%%%%%%

Lastly, we investigate the rate of the charge separation influenced by all of the vibrational modes ($\omega_\xi < 1200\,\cm$) in chlorophyll and pheophytin molecules given in \fig~\ref{fig:HR-factors-CS}. Figure~\ref{fig:rate-allmodes} presents the rate as a function of the driving force $-\Delta G^\circ$ for various values of the environmental reorganization energy $\lambda_{\A\D}$.
The panels on the left (a--c) present the results for the physiological temperature $T=300\,\K$, whereas those on the right (d--f) present those for the cryogenic temperature $T=77\,\K$.
The calculations were performed with \eq~\ref{eq:employed-eq}, of which the applicability is verified in \fig~\ref{fig:rate-1mode}. 
Although the rate enhancement caused by individual vibrational modes is small as in \fig~\ref{fig:rate-1mode}, these small contributions accumulate to make a large difference, in particular in the inverted region, $-\Delta G^\circ > \lambda_{\A\D}$. 
On the other hand, the rate is observed to decrease in the normal region, $-\Delta G^\circ < \lambda_{\A\D}$. An electron transfer reaction occurring in the normal region typically involves only vibrational ground states, and the coupling between the donor and acceptor states is described with the Franck-Condon factors of the involved vibrational modes as $ V_{\D\A} \to V_{\D\A} \prod_{\xi=1}^N \e^{-S_\xi/2}$. See \eq~\ref{eq:Jortner-Bixon-2}.
This reduction in the interstate coupling is responsible for the suppression of the charge transfer rate in the normal region.
The vibrational contributions not only cause the rate enhancement in the inverted region but also the rate suppression in the normal region. 
Although contributions of the low-frequency vibrational modes ($\omega_\xi<150\,\cm$) were excluded in the presented results, we verified that inclusion of such low-frequency modes further suppressed the rate in the normal region and hardly varied the rate in the inverted region.
Consequently, the time constant of the charge separation is maintained to be sub-picosecond, which is consistent with the experimental results,\cite{Groot:2005tk, Holzwarth:2006wt, Raszewski:2008go} over a wide range of the driving force $-\Delta G^\circ$ and the environmental reorganization energy $\lambda_{\A\D}$. 
In general, static disorder in systems with larger reorganization energy is stronger, and Gelzinis et al. obtained the large static disorder ($\sim 550\,\cm$) by analyzing multiple optical spectra of the PSII RC.\cite{Gelzinis:2017cz}
In this respect, \fig~\ref{fig:rate-allmodes} indicates the inherent robustness of the rate of the primary charge separation in the PSII RC against the disorder in the involved electronic energies as well as in the environmental reorganization energy.
Moreover, the insensitivity to temperature variations ($77-300\,\K$) is consistent with the experimental results of the sub-picosecond transient absorption spectroscopy by Groot et al.\cite{Groot:1997tj}

In conclusion, we determined that the electronic-vibrational resonance of individual vibrational modes plays a minor role in promoting the charge separation process starting from the accessory Chl in the protein environment, contrary to the discussion in recent publications.\cite{Romero:2014jm,Fuller:2014iz} 
The free energy of activation required for the charge separation to proceed is no larger than the thermal energy at physiological temperatures over a wide range of the environmental reorganization energy and the driving force. Hence, the charge separation can take place in a facile fashion without the assistance of the high-frequency vibrational modes.
However, our examination of the impacts of all of the intramolecular vibrations revealed that these small contributions add up to ultimately have a large influence on the charge separation rate, namely a decrease in the normal region and an increase in the inverted region.
This change enables the charge separation rate of sub-picoseconds to be almost independent of the driving force, environmental reorganization energy, and temperature variations.
We suggest that intramolecular vibration may represent an important design principle that enables the high quantum efficiency of charge separation in oxygenic photosynthesis in the sense that it complements the robustness of the charge separation in the photosystem II reaction center against the inherently large static disorder of the involved electronic energies.
In this work, we focused only on the charge separation starting from the accessory $\rm Chl_{D1}^\ast$. As was discussed by Romero et al.\cite{Romero:2014jm,Novoderezhkin:2017jta} and Fuller et al.,\cite{Fuller:2014iz} however, the charge separation from the special pair $(\rm P_{D1}P_{D2})^\ast$ cannot be easily discounted. Detailed analyses on the special pair are left for future studies.

% Appendix %%%%%%%%%%%
\section*{Methods}

\paragraph*{Quantum chemical calculations.}
Higashi et al.\cite{Higashi:2014ga} demonstrated that time-dependent density functional theory (TDDFT) with a range-separated hybrid functional, CAM-B3LYP \cite{Yanai:2004ia} reproduced the lowest singlet excitation energy of bacteriochlorophyll\,{\it a} (BChl{\it a}) in various organic solvents.  This functional describes the long-range correction to the DFT exchange functional scheme by mixing the DFT short-range term with the Hartree-Fock exchange-integral term that expresses long-range orbital-orbital-exchange interactions, depending on the electronic distance.\cite{Yanai:2004ia} The mixing is characterized by the parameter $\mu$. In the original CAM-B3LYP, the value of $\mu$ is set to be 0.33; however, Higashi et al. found that the $\Qy$ transition energy of BChl{\it a} in organic solvents was best reproduced by employing $\mu = 0.2$.\cite{Higashi:2014ga} With this value, they also succeeded in reproducing the $\Qy$ transition energies of the seven BChl{\it a} molecules embedded in the Fenna-Matthews-Olson protein of green sulfur photosynthetic bacteria.\cite{Higashi:2016fi} Along this line, Saito et al.\cite{Saito:2018ig} demonstrated that TDDFT calculations with the CAM-B3LYP parameter $\mu = 0.14$ reproduced the observed $\Qy$ transition energies of Chl{\it a} and Chl{\it b} in organic solvents, suggesting $\mu = 0.14$ would be applicable to Chl{\it a}/Chl{\it b}-binding proteins such as PSII and light harvesting complex II (LHCII). In this study, therefore, we perform TDDFT calculations based on the CAM-B3LYP functional with $\mu=0.14$ for geometry optimization of Chl{\it a} and Pheo{\it a} molecules. We used the 6-31G(d) basis set as in the previous study.\cite{Saito:2018ig} The computational cost was reduced by replacing the long alkyl chains in Chl{\it a} and Pheo{\it a} molecules with methyl groups.\cite{Higashi:2014ga}

\paragraph*{Numerically accurate quantum dynamics calculations.}
By utilizing the definition of the solvation coordinate $\X_{\A\D}$, the Hamiltonian in \eq~\ref{eq:primary-CS-hamiltonian} is recast into the system-plus-environment form,
\begin{align}
	\H = \H_\sys + \H_\sysenv + \H_\env
\end{align}
with $\H_\sys = \varepsilon \lvert \D \rangle\langle \D \rvert  + ( \varepsilon + \hbar\Omega_{\A\D} ) \lvert \A \rangle\langle \A \rvert  + V_{\D\A}(\lvert \D \rangle\langle \A \rvert  + \lvert \A \rangle\langle \D \rvert)$, $\H_\sysenv= \X_{\A\D} \lvert \A \rangle\langle \A \rvert$, and $\H_\env=\H_\D$.
Here, $\varepsilon$ is an arbitrary c-number. 
It is noted that the form of the system-environment interaction Hamiltonian differs from that for electronic energy transfer in condensed phases.\cite{Romero:2014jm,Fuller:2014iz,SchlauCohen:2011cw}
An adequate description of the charge separation dynamics is provided with the reduced density operator, \ie the partial trace of the density operator of the total system over the environmental and nuclear DOFs.
When the environmentally induced fluctuations are described as Gaussian processes and the relevant nuclear dynamics are harmonic, the time evolution of the reduced density operator for the Hamiltonian can be integrated in a numerically accurate fashion by using the so-called hierarchical equations of motion approach.\cite{Tanimura:2006ga} The charge separation rate is evaluated in accordance with \refer~\citenum{Ishizaki:2009jg}.

\paragraph*{The Marcus and Jortner-Bixon formulae.}
In \eq~\ref{eq:cummurant-rate}, we assume the following:
[1] the high-temperature limit for the environmental DOFs, $\coth(\beta\hbar\omega/2)\simeq 2/\beta\hbar\omega$,
[2]  the short-time approximation for the environmental DOFs, $\cos\omega t \simeq 1 - \omega^2t^2/2$ and $\sin\omega t \simeq \omega t$,
[3] the low-temperature limit for the vibrational DOFs, $\coth(\beta\hbar\omega/2)\simeq 1$, 
and
[4] infinitely slow vibrational relaxation and correspondingly the $\delta$-function form for the vibrational spectral density, $J_{\vib}(\omega)=\sum_{\xi=1}^N \pi\hbar S_\xi \omega_\xi^2 [\delta(\omega-\omega_\xi) - \delta(\omega+\omega_\xi)]$.
Under these assumptions, \eq~\ref{eq:cummurant-rate} leads to the multimodal expression of the Jortner-Bixon formula,\cite{Jortner:1988jr}
\begin{align}
	k^{(2)}
	=
	\sqrt{\frac{\pi}{\hbar^2 \lambda_{\A \D}  k_{\rm B}T}}
	\sum_{n_1=0}^\infty\dots\sum_{n_N=0}^\infty
	\kappa(\{n_\xi\})
	\label{eq:Jortner-Bixon-1}
\end{align}
with
\begin{align}
	\kappa(\{n_\xi\})
	=
	V_{\D \A}^2
	\left(
	\prod_{\xi=1}^N 
	\e^{-S_\xi} \frac{{S_\xi}^{n_\xi}}{n_\xi!}
	\right)
	\e^{-{\Delta G^\ast(\{n_\xi\})}/{k_{\rm B}T}}.	
	\label{eq:Jortner-Bixon-2}
\end{align}
In the above, non-negative integer $n_\xi$ stands for a vibrational quantum number of the $\xi$th high-frequency mode that satisfies $\hbar\omega_\xi > k_{\rm B}T$. 
The corresponding free energy of activation $\Delta G^\ast(\{n_\xi\})$ is expressed as
$
	\Delta G^\ast(\{n_\xi\})
	=
	{(\Delta G^\circ + \lambda_{\A \D}+\hbar\sum_{\xi=1}^N  n_\xi \omega_\xi)^2}/{4\lambda_{\A \D}}.
$
In the absence of the vibrational contributions ($n_\xi\to 0$ and $S_\xi \to 0$), \eq~\ref{eq:Jortner-Bixon-1} yields the Marcus formula.\cite{Marcus:1993kx}

% Acknowledgements %%%%%%%%%%%
\begin{acknowledgements}
The authors are grateful to Professor~Jeffrey~Reimers for providing the code of his DUSHIN program.
Numerical calculations were partly performed at the Research Center for Computational Science, Okazaki Research Facilities, National Institutes of Natural Sciences. 
This work was supported by JSPS KAKENHI Grant Numbers 16KT0165, 17H02946, 17K05757, and 18H01937 as well as JSPS KAKENHI Grant Numbers 17H06437 in the Innovative Area ``Innovations for Light-Energy Conversion (I$^4$LEC)'' and 18H04657 in the Innovative Area ``Hybrid Catalysis.''
\end{acknowledgements}

% Bibliography %%%%%%%%%%%%%%%
%

\end{document}